\documentstyle[multicol,aps]{revtex}
\renewcommand{\narrowtext}{\begin{multicols}{2}
\global\columnwidth20.5pc\noindent}
\renewcommand{\widetext}{\end{multicols}
\global\columnwidth42.5pc}
\begin{document}
\title{Combination of Ferromagnetic and Antiferromagnetic Features in
       Heisenberg Ferrimagnets}
\author{Shoji Yamamoto}
\address{Department of Physics, Faculty of Science, Okayama University,\\
         Tsushima, Okayama 700-8530, Japan}
\author{Takahiro Fukui}
\address{Institut f\"ur Theoretische Physik, Universit\"at zu K\"oln,\\
         Z\"ulpicher Strasse 77, 50937 K\"oln, Germany}
\author{Klaus Maisinger and Ulrich Schollw\"ock}
\address{Sektion Physik, Ludwig-Maximilians-Universit\"at M\"unchen,\\
         Theresienstrasse 37, 80333 Munich, Germany}
\date{Received 19 June 1998}
\maketitle
\begin{abstract}
   We investigate the thermodynamic properties of Heisenberg ferrimagnetic
mixed-spin chains both numerically and analytically with particular
emphasis on the combination of ferromagnetic and antiferromagnetic
features.
Employing a new density-matrix renormalization-group technique as well
as a quantum Monte Carlo method, we reveal the overall thermal
behavior:
At very low temperatures, the specific heat and the magnetic
susceptibility times temperature behave like $T^{1/2}$ and
$T^{-1}$, respectively, whereas at intermediate temperatures, they
exhibit a Schottky-like peak and a minimum, respectively.
Developing the modified spin-wave theory, we complement the numerical
findings and give a precise estimate of the low-temperature behavior.
\end{abstract}

\narrowtext
\section{Introduction}\label{S:I}

   Low-dimensional quantum magnets with two kinds of
antiferromagnetically exchange-coupled centers have been attracting
much current interest.
Several authors \cite{Vega99,Alad29,Fuji81,Dorf31} constructed
integrable Hamiltonians and extracted suggestive critical
phenomena from them.
Alternating-spin antiferromagnets with singlet ground states
stimulate us to study again the nontrivial gap problem \cite{Hald64}.
Employing the nonlinear-$\sigma$-model technique, various mixed-spin
systems such as linear chains \cite{Fuku09,Taka55} and ladders
\cite{Fuku98} have systematically been studied with particular
emphasis on the competition between massive and massless  phases.

   In this context a remarkable attention has been directed to
ferrimagnetic mixed-spin chains
\cite
{Alca67,Pati94,Kole36,Breh21,Nigg31,Yama09,Ono76,Kura29,Yama91,Ivan14},
which are the subject in the present article.
Since we may expect gapless excitations from magnetic ground states,
it is less interesting there whether the system is massless or
massive.
Using a field-theoretical argument, Alcaraz and Malvezzi
\cite{Alca67} predicted that mixed-spin isotropic Heisenberg
ferrimagnets should exhibit quadratic dispersion relations.
Their prediction was numerically verified and the quadratic dispersion
was explicitly visualized \cite{Yama09}.
Thus quantum ferrimagnets are expected to behave like ferromagnets
at low temperatures.
On the other hand, conventional spin-wave calculations
\cite{Pati94,Breh21} and a perturbation approach \cite{Yama09} from
the decoupled-dimer limit suggest that quantum ferrimagnets should
exhibit nontrivial gapped excitations as well, which have the effect of
enhancing the ground-state magnetization and are therefore of
antiferromagnetic nature.
A quantum Monte Carlo (QMC) technique and an exact-diagonalization
method \cite{Yama09} actually observed the two distinct low-lying
excitations and showed that the mixed nature remains
unchanged as long as the model is isotropic.

   Motivated by the revealed low-energy structure, two of the
present authors \cite{Yama91} have investigated the thermodynamic
properties of Heisenberg ferrimagnetic spin chains focusing on the idea of
coexisting ferromagnetic and antiferromagnetic aspects.
On the one hand, a QMC method was employed to find the overall
thermal behavior.
On the other hand, a modified spin-wave (MSW) theory was introduced
to clarify how the two distinct excitations contribute to the
thermodynamics.
The present article is a developed richer version of the preceding
work.
In order to inquire further into the low-temperature behavior, we
here employ additional numerical tools.
A quantum transfer-matrix (QTM) method allows us to observe how the
ferromagnetic character grows with the increase of the system size.
A brand-new density-matrix renormalization-group (DMRG) technique
helps us to go down into the low-temperature region that was never
attained before.
Furthermore, taking account of certain interactions between spin
waves, we refine the MSW theory so as to bring a more accurate
description of the thermal quantities at low temperatures.
Although it is rather hard, even with interacting spin waves,
to obtain a quantitative description of the overall thermal behavior,
a grand-canonical approach to ferrimagnets in terms of spin waves is
suggestive enough and interesting in itself.
We fully argue how we should {\it modify} the conventional spin-wave
theory in an attempt to construct the thermodynamics of quantum
ferrimagnets.

   We consider two kinds of spins $S$ and $s$ alternating on a ring
with antiferromagnetic exchange coupling between nearest neighbors,
as described by the Hamiltonian
\begin{equation}
   {\cal H}
      =J\sum_{j=1}^N
        \left(
         \mbox{\boldmath$S$}_{j} \cdot \mbox{\boldmath$s$}_{j}
        +\delta
         \mbox{\boldmath$s$}_{j} \cdot \mbox{\boldmath$S$}_{j+1}
        \right)
      -g\mu_{\rm B} H\sum_{j=1}^N(S_j^z+s_j^z)\,,
   \label{E:H}
\end{equation}
where $N$ denotes the number of unit cells, $\delta$ represents a
bond alternation, $\mu_B$ is the Bohr magneton, and we have set the
$g$ factors of both spins $S$ and $s$ equal to $g$.
We assume that $S>s$, which remains general enough to describe
alternating-spin ferrimagnets.
The Lieb-Mattis theorem \cite{Lieb49} shows that the Hamiltonian
(\ref{E:H}), unless a field applied, has $[2(S-s)N+1]$-fold
degenerate ground states.
The ferromagnetic and the antiferromagnetic excitations, which
lie in the subspaces of $M<(S-s)N$ and $M>(S-s)N$, respectively,
indeed show a quadratic dispersion and a gapped spectrum
\cite{Yama09}, where $M=\sum_{j=1}^N(S_j^z+s_j^z)$ is the total
magnetization.
The antiferromagnetic gap, namely, the gap between the ground state
and the lowest excitation to $M=(S-s)N+1$, was estimated to be
$1.75914(1)J$ in the thermodynamic limit.
The correlation length of the system is so small that it is barely
of the length of the unit cell \cite{Pati94,Breh21} even at the
Heisenberg point $\delta=1$.

\section{Numerical Study}\label{S:NS}

   In this section, using several numerical tools, we calculate the
specific heat and the magnetic susceptibility at zero field.
The spin-wave \cite{Pati94,Breh21} and the perturbation \cite{Yama09}
calculations suggest that the low-temperature properties of the model
are qualitatively the same regardless of the values of $S$ and $s$ as
long as they differ from each other.
Alcaraz and Malvezzi \cite{Alca67} performed finite-size calculations
combined with a scaling analysis in the cases of $(S,s)=(1,1/2)$ and
$(S,s)=(3/2,1/2)$ and indeed concluded that the appearance of
quadratic dispersion relations should be expected for arbitrary
isotropic mixed-spin chains showing ferrimagnetism instead of
antiferromagnetism.
Thus we restrict our numerical investigations to the case of
$(S,s)=(1,1/2)$.

\subsection{Procedure}

   We employ the QMC method based on the Suzuki-Trotter decomposition
\cite{Suzu54} of the checkerboard type \cite{Hirs33}.
The partition function $Z={\rm Tr}[{\rm e}^{-\beta{\cal H}}]$
is approximately decomposed as 
\begin{equation}
   Z\simeq
     \left[
      \left(
       \prod_{i=1,3,\cdots}{\rm e}^{-\beta h_i/n}
       \prod_{i=2,4,\cdots}{\rm e}^{-\beta h_i/n}
      \right)^n
     \right]\,,
   \label{E:STdec}
\end{equation}
where $n$ is a Trotter number, $\beta=(k_{\rm B}T)^{-1}$ with the
Boltzmann constant $k_{\rm B}$, and
\begin{equation}
   \left.
   \begin{array}{l}
   {\displaystyle
   h_{2j-1}
     =J
      \mbox{\boldmath$S$}_{j} \cdot \mbox{\boldmath$s$}_{j}
     -\frac{1}{2}g\mu_{\rm B}H(S_j^z+s_j^z)\,,
   }
   \\
   {\displaystyle
   h_{2j}
     =J\delta
      \mbox{\boldmath$s$}_{j} \cdot \mbox{\boldmath$S$}_{j+1}
     -\frac{1}{2}g\mu_{\rm B}H(s_j^z+S_{j+1}^z)\,.
   }
   \end{array}
   \right.
   \label{E:localH}
\end{equation}
In order to accelerate the thermodynamic calculation and refine its
accuracy, we make use of the improved algorithm in
Ref.\ \onlinecite{Yama49}.
Calculations are carried out at several values of $n$ and $N$ and
they are extrapolated to the limit of $n\rightarrow\infty$ and
$N\rightarrow\infty$.
Although we have treated chains of $N=24$, $N=32$, and $N=48$, the
size dependence of the thermal quantities is so weak that we find
no difference beyond the numerical uncertainty even
between the calculations at $N=24$ and $N=32$ except for very low
temperatures.
Both the specific heat $C$ and the magnetic susceptibility $\chi$ are
directly evaluated through formulae \cite{Yama64}
\begin{eqnarray}
   &&
   C=k_{\rm B}\beta^2
     \left(
      \langle Q_2  \rangle
     +\langle Q_1^2\rangle
     -\langle Q_1  \rangle^2
     \right)\,,
   \label{E:QMC-C} \\
   &&
   \chi=g^2\mu_{\rm B}^2\beta
     \left(
      \langle M_2\rangle
     -\langle M_1\rangle^2
     \right)\,,
   \label{E:QMC-S}
\end{eqnarray}
with
\begin{equation}
   \left.
   \begin{array}{l}
   {\displaystyle
   Q_1=\sum_{i,m}
       \biggl[
        \frac{\partial\rho_i^{(m)}}{\partial\beta}
        /\rho_i^{(m)}
       \biggr]\,,
   }
   \\
   {\displaystyle
   Q_2=\sum_{i,m}
       \biggl[
        \frac{\partial^2\rho_i^{(m)}}{\partial\beta^2}
        /\rho_i^{(m)}
       -\Bigl(
         \frac{\partial\rho_i^{(m)}}{\partial\beta}
         /\rho_i^{(m)}
        \Bigr)^2
       \biggr]\,,
   }
   \end{array}
   \right.
   \label{E:QMC-C2}
\end{equation}
\begin{equation}
   \left.
   \begin{array}{l}
   {\displaystyle
   M_1=\frac{1}{2n}\sum_{i,m}
       \sigma_i^{(m)}\,,
   }
   \\
   {\displaystyle
   M_2=\frac{1}{2n}\sum_{m}
       \Bigl(
        \sum_i\sigma_i^{(m)}
       \Bigr)^2\,.
   }
   \end{array}
   \right.
   \label{E:QMC-S2}
\end{equation}
where $\langle A\rangle$ denotes the thermal average of $A$ at a
given temperature $\beta^{-1}$,
$\sigma_i^{(2l-1)}$ and $\sigma_i^{(2l)}$ are Ising spins on the site
$i$ at the $l$th Trotter layer, and the local Boltzmann factors
$\rho_{2j-1}^{(2l-1)}$ and $\rho_{2j}^{(2l)}$ are defined in terms of
the eigenstates of $S_i^z$ and $s_i^z$ as
\begin{equation}
   \rho_i^{(m)}
     =\langle
       \sigma_i^{(m)},\sigma_{i+1}^{(m)}
       \vert{\rm e}^{-\beta h_i/n}\vert
       \sigma_i^{(m+1)},\sigma_{i+1}^{(m+1)}
      \rangle \,.
   \label{E:BF}
\end{equation}

   As far as we replace the trace by an importance sampling, it is
hardly feasible to take grandcanonical averages at very low
temperatures.
In an attempt to avoid the difficulty, we may integrate out
\cite{Bets29,Suzu57} the $(1+1)$-dimensional Ising system but then
have to give up either low temperatures or long chains.
We can in principle reach an arbitrary low temperature at the expense of
the system size and such an attempt is to a certain extent fruitful
in the present system whose correlation length is very
small.
Actually, constructing transfer matrices in the chain direction,
we will later show QTM calculations of the specific heat.
Although the QTM calculations themselves are so accurate as to be
regarded as exact, numerical differentiations of them may introduce 
errors.
Here we directly calculate the internal energy, rather than the
partition function, and numerically differentiate it once.
We keep the final results highly accurate by taking raw data at
regular intervals of $k_{\rm B}T/J=0.01$.

   On the other hand, constructing transfer matrices in the Trotter
direction, we can partly reveal the thermal behavior of an infinite
chain \cite{Bets19}.
However, the exponential growth of the matrix size with the increase
of $n$ makes our access to low temperatures unfeasible.
In order to overcome the difficulty, we introduce the DMRG technique
into our investigations.
Based on White's original idea \cite{Whit63} developed at $T=0$,
we apply it to the renormalization of transfer matrices
\cite{Nish98,Burs83} instead of Hamiltonians.
This extension of the DMRG method has recently been applied to the
thermodynamics of several low-dimensional magnets
\cite{Shib21,Wang61,Mais22} and indeed brought us plenty of results
at very low temperatures.

   The core idea of the so-called transfer-matrix DMRG method can be
summarized as reaching large $n$ by reducing the size of the transfer
matrix with the use of the DMRG algorithm, which is highly successful
in obtaining effective low-energy Hamiltonians.
We introduce a stepwidth $\beta_0$ and go successively down to lower
temperatures $\beta^{-1}=(n\beta_0)^{-1}$ increasing $n$ linearly.
The transfer-matrix DMRG calculation is technically different from
the original procedure in that we have to handle nonsymmetric
density matrices there.
In order to avoid the exponential growth of the matrix size, we keep
the number of states in the density matrix constant at a predetermined
number $m$ throughout the calculation.
At each iteration $n\rightarrow n+1$, we first produce the transfer
matrix at the step $n+1$ by making a matrix product of the transfer
matrix at the step $n$ and the local Boltzmann factor (\ref{E:BF})
with $\beta=n\beta_0$.
Next we unite the two transfer matrices at the step $n+1$ into an
augmented matrix of double width in the Trotter direction.
We then calculate the left and right eigenvectors of the enlarged
matrix with the maximum eigenvalue and construct a density matrix
from them.
It is the eigenstates corresponding to the $m$ largest eigenvalues
of the density matrix that should be retained to give an approximate
description of the transfer matrix at the step $n+1$.
The thus-obtained transfer matrix allows us to calculate the free
energy at each temperature $\beta^{-1}=(n\beta_0)^{-1}$.
We here carry out the direct estimation \cite{Burs83} of the internal
energy and the magnetization, which strongly reduces numerical errors
in the final results, the specific heat and the magnetic
susceptibility.

   Obviously the two controllable parameters $\beta_0$ and $m$ decide
the precision of the transfer-matrix DMRG calculation.
The Trotter decomposition is refined as $\beta_0\rightarrow 0$,
whereas the loss of information is reduced as $m\rightarrow\infty$.
At high temperatures, we are compelled to work with small Trotter
numbers, while few states are discarded.
At low temperatures, Trotter numbers are large enough, whereas many
states are lost due to the large number of iterations.
Thus, the convergence of the calculation predominantly depends on
$\beta_0$ at high temperatures, while on $m$ at low temperatures, as
was actually observed \cite{Mais22}.

   As we have conclusive QMC results at high
and intermediate temperatures, we are particularly
interested in low-temperature findings
from the transfer-matrix DMRG calculation.
We therefore invest computational resources mainly in augmenting $m$.
Setting $\beta_0J$ and $m$ to $0.2$ and $128$, respectively, we have
found that at intermediate temperatures the specific heat is
somewhat overestimated, whereas at low temperatures the precision
almost reaches three digits.
Calculations at $m=80$, $m=96$ and $m=128$ fully converge at
$T\agt 0.05$ and almost converge at $0.04\alt T\alt 0.05$.
We were not able to go further down below successfully.
This is mainly due to the macroscopically degenerate ground states of
ferrimagnets, which imply that a huge number of states must be
retained in our access $T\rightarrow 0$.

\subsection{Results}

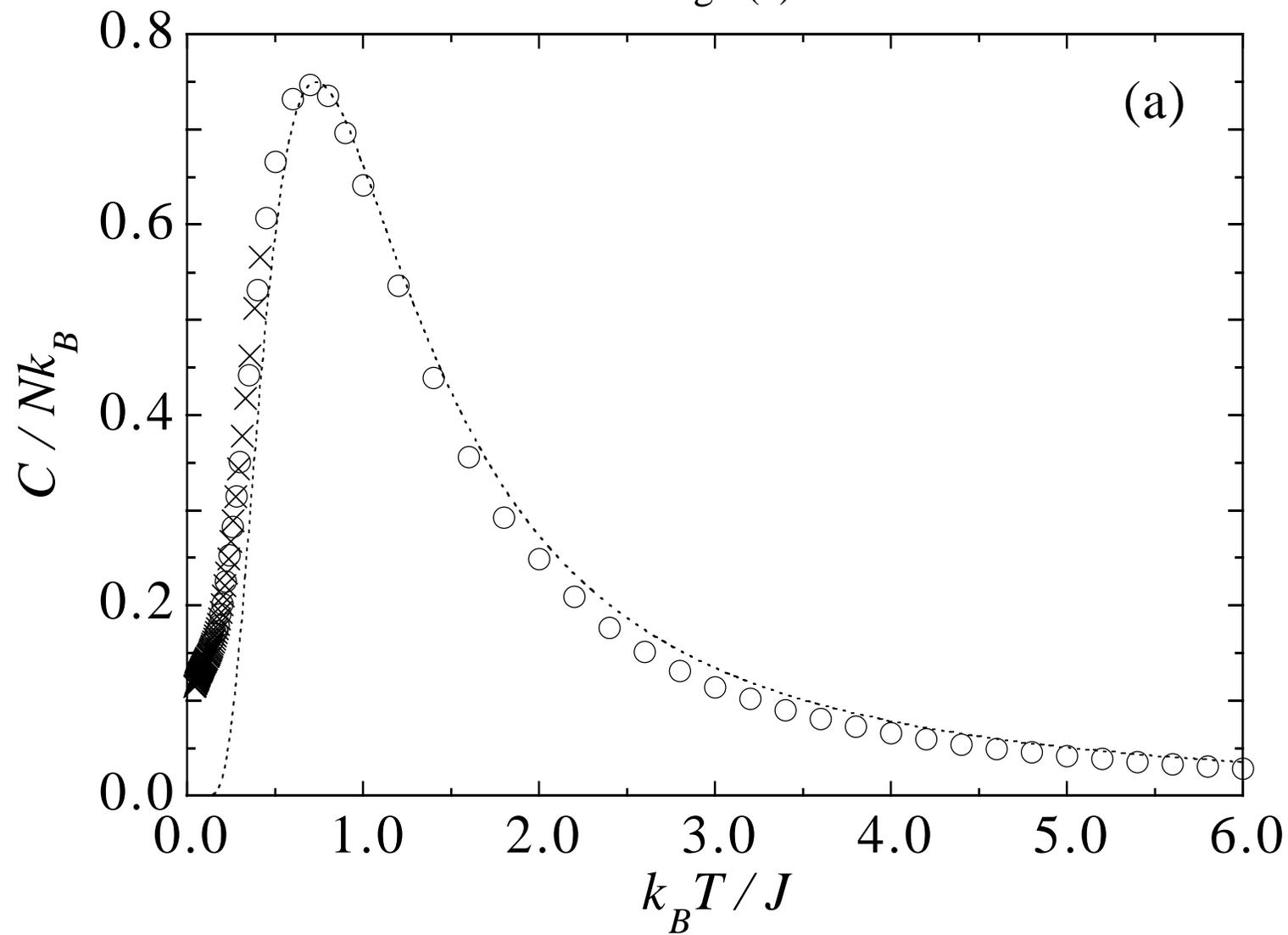
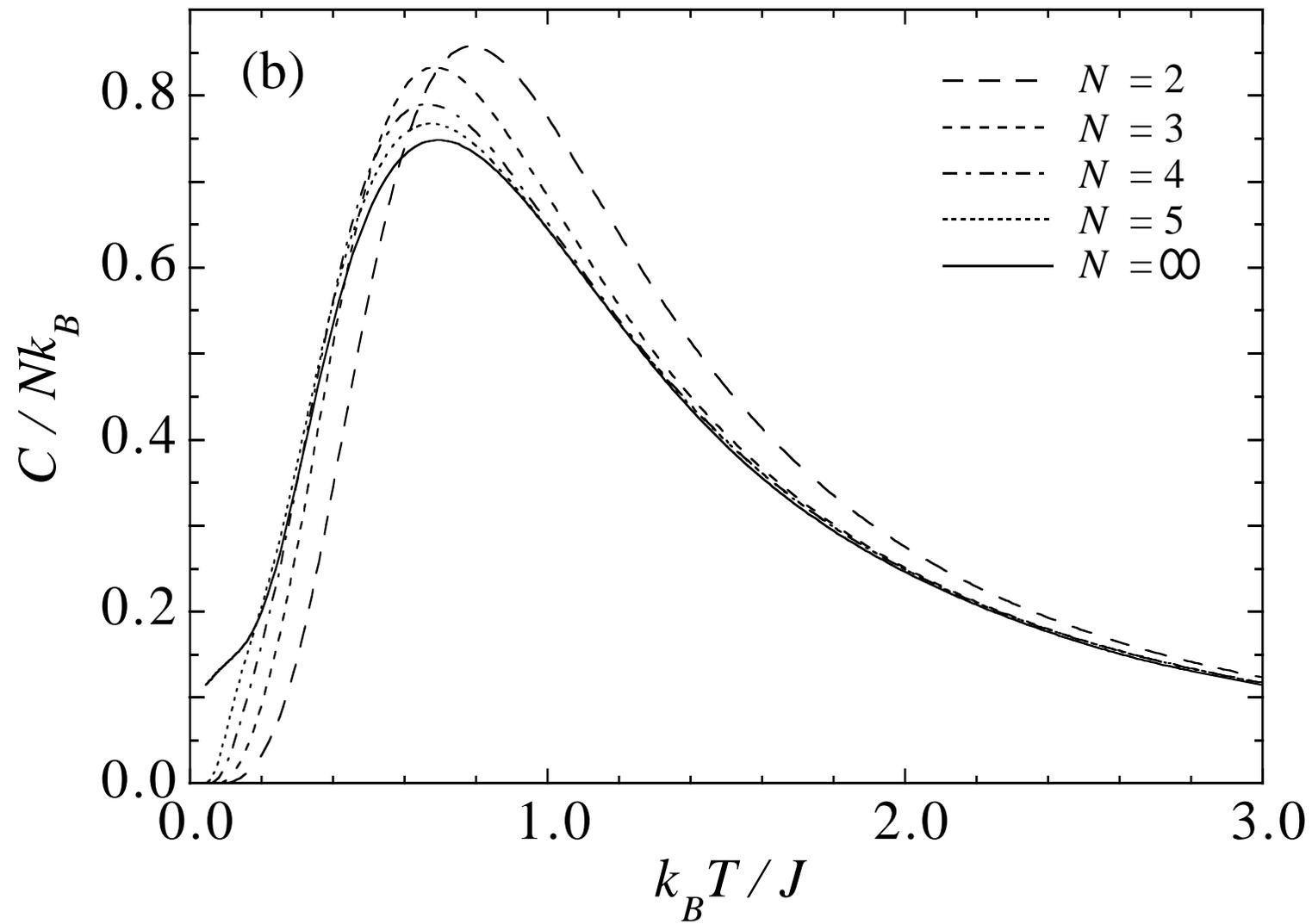
\begin{figure}
\caption{Temperature dependence of the specific heat per unit cell:
         (a) QMC findings ($\bigcirc$) and DMRG results ($\times$)
         at $N\rightarrow\infty$.
         The dotted line represents the Schottky-type specific heat
         (2.9).
         The numerical uncertainty is smaller than the symbol size.
         (b) QTM calculations at various values of $N$, where we show
         the $N\rightarrow\infty$ curve as well, which is obtained by
         interconnecting the QMC findings at $k_{\rm B}T/J\geq 0.4$
         and the DMRG results at $k_{\rm B}T/J\leq 0.4$.}
\label{F:Num-C}
\end{figure}

   We show in Fig. \ref{F:Num-C} the temperature dependence of the
specific heat of the $(S,s)=(1,1/2)$ Heisenberg ferrimagnetic spin
chain.
We find in Fig. \ref{F:Num-C}(a) a good agreement between the QMC
and DMRG calculations.
At intermediate temperatures, the specific heat
exhibits a Schottky-like anomaly typical of antiferromagnets.
Let us consider the model at $\delta=0$ in an attempt to clarify the
origin of this characteristic peak.
Now the model is decoupled into dimers and the specific heat per
unit cell is simply obtained from an isolated dimer, which is a
pure two-level system with doubly degenerate ground states and
fourfold degenerate excited states separated above by an energy gap
${\mit\Delta}=3J/2$.
The specific heat is given by
\begin{equation}
   \frac{C}{Nk_{\rm B}}
     =\frac{r(\beta{\mit\Delta})^2{\rm e}^{\beta{\mit\Delta}}}
           {({\rm e}^{\beta{\mit\Delta}}+r)^2} \,,
   \label{E:Schottky}
\end{equation}
where $r$ is the ratio of the degeneracy of the excited states to that
of the ground states.
Moving away from the decoupled-dimer point, the low-lying states come
to exhibit dispersion and macroscopic degeneracy.
At the Heisenberg point $\delta=1$, 
for $N$ elementary cells, the ground state is
$(N+1)$-fold degenerate, whereas the lowest antiferromagnetic
excited state is $(N+3)$-fold degenerate \cite{Lieb49}.
We note that the isolated dimer may be regarded as the Heisenberg
chain of $N=1$.
Now we make an attempt to fit the intermediate-temperature
behavior to the curve (\ref{E:Schottky}) allowing only a single
additional adjustable parameter $A$:
\begin{equation}
   \frac{C}{Nk_{\rm B}}
     =\frac{Ar(\beta{\mit\Delta})^2{\rm e}^{\beta{\mit\Delta}}}
           {({\rm e}^{\beta{\mit\Delta}}+r)^2} \,.
   \label{E:Schottky2}
\end{equation}
Here we replace $r$ and ${\mit\Delta}$ by their values at the
Heisenberg point, the degeneracy ratio of the lowest
antiferromagnetic excited states to the ground states and the gap
between them, respectively:
\begin{equation}
   r=\lim_{N\rightarrow\infty}\frac{N+3}{N+1}=1\,,\ \ 
   {\mit\Delta}=1.759J\,.
   \label{E:fit}
\end{equation}
We obtain a fine fit with $A=1.7$, which is also shown in Fig.
\ref{F:Num-C}(a).
We stress that the Schottky-type specific heat (\ref{E:Schottky2})
not only fits the peak but also well describes the high-temperature
decay.
Equation (\ref{E:Schottky2}) combined with the condition
(\ref{E:fit}) gives an asymptotic high-temperature behavior
$1.3(\beta J)^2$, which is well consistent with the high-temperature
series-expansion result $(\beta J)^2$ \cite{Yama91}.
Thus we find that the temperature dependence of the specific heat
straightforwardly reflects the antiferromagnetic gap and therefore
the model can be treated well as a two-level system unless it is at low
enough temperatures $k_{\rm B}T\ll{\mit\Delta}\simeq 1.759J$.

   On the other hand, the DMRG results fully allow us to guess the
$T^{1/2}$ asymptotic behavior characteristic of ferromagnets.
This ferromagnetic feature grows as we move toward the Heisenberg
point from the decoupled-dimer limit or alternatively as the system
size increases.
From this point of view, we observe the chain-length dependence of
the specific heat in Fig. \ref{F:Num-C}(b).
The naive QTM method is most useful in short chains at low
temperatures.
We find that the antiferromagnetic feature is rather settled even
in short chains, whereas the growth of the ferromagnetic feature is
relatively slow.
The elementary excitations of ferromagnetic nature are predominantly
related to spin $1$'s, while those of antiferromagnetic nature
originate from interactions between the two kinds of spins
\cite{Yama09}.
These excitations may be identified with the local ones within a unit
cell in the vicinity of the decoupled-dimer limit, as shown in Fig.
\ref{F:dimer}.
Figure \ref{F:Num-C}(b) suggests that the delocalization effect is
more essential to the appearance of the ferromagnetic feature.
Actually the ferromagnetic excitations constitute a wider band than
the antiferromagnetic ones \cite{Yama09}.

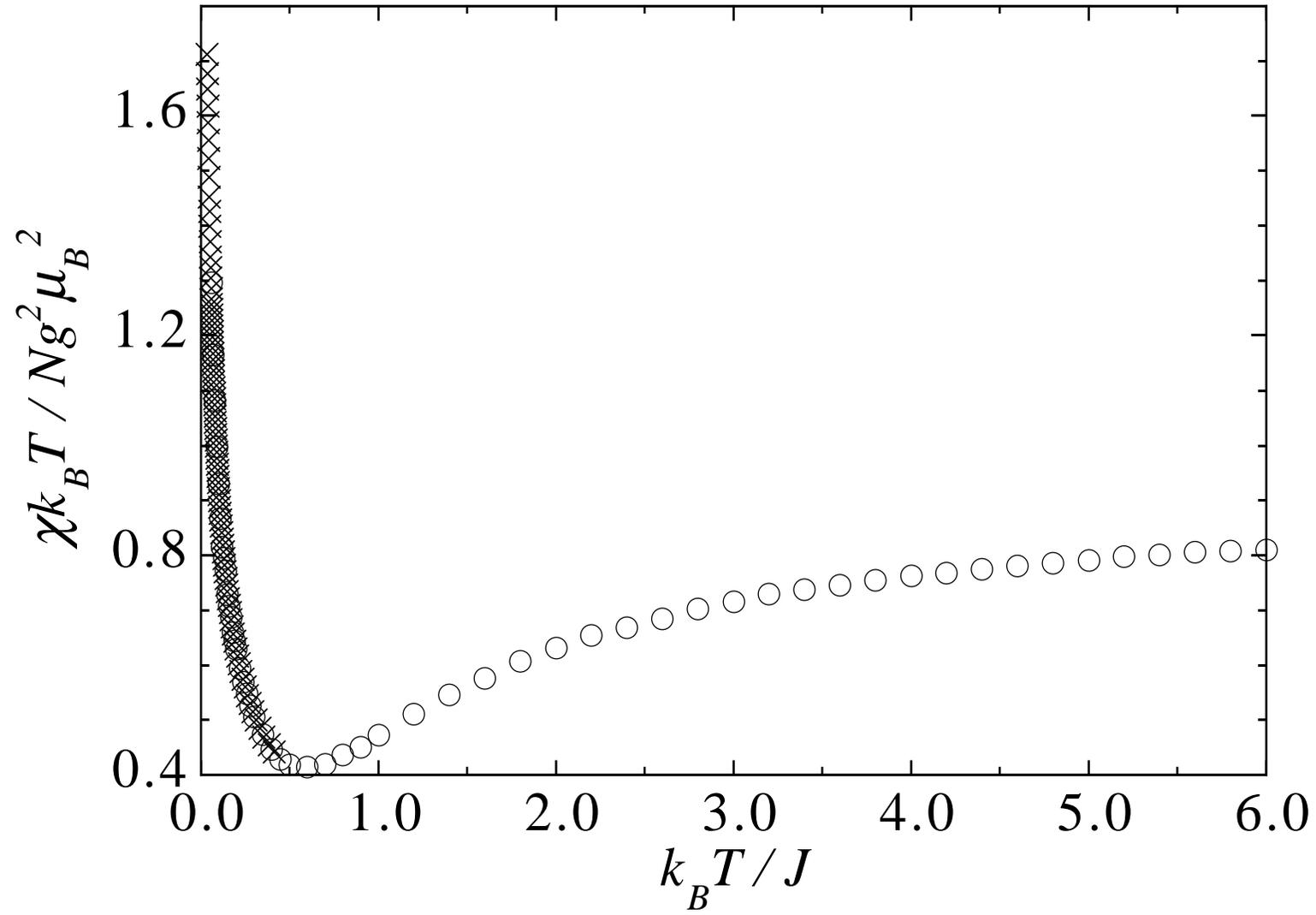
\begin{figure}
\caption{Temperature dependence of the magnetic susceptibility times
         temperature per unit cell:
         QMC findings ($\bigcirc$) and DMRG results ($\times$)
         at $N\rightarrow\infty$.
         The numerical uncertainty is smaller than the symbol size.}
\label{F:Num-S}
\end{figure}

   We show in Fig. \ref{F:Num-S} the temperature dependence of the
magnetic susceptibility times temperature of the
$(S,s)=(1,1/2)$ Heisenberg ferrimagnetic spin chain.
We find again that the QMC and DMRG calculations are in excellent
agreement.
The product $\chi T$ diverges as $T^{-1}$ at low temperatures, while it
approaches $[S(S+1)+s(s+1)]/3=11/12$ at high temperatures.
The low-temperature divergence is reminiscent of the ferromagnetic
susceptibility \cite{Taka08}.
We further note that $\chi T$ shows a minimum while shifting from the
quantum ferromagnetic behavior to the classical paramagnetic
behavior.
Generally $\chi T$ is a monotonically decreasing function in
ferromagnets, while a monotonically increasing function in
antiferromagnets \cite{Bonn40}.
The susceptibilities of gapped antiferromagnets such as Haldane
systems vanish exponentially \cite{Yama49,Yama64,Yama28}.
Hence the temperature dependence we are observing may be regarded as
a ferromagnetic-to-antiferromagnetic crossover.
Actually the minimum of $\chi T$ appears at a temperature around
which the specific heat shows the Schottky-like peak.

   In sum, a combination of ferromagnetic and antiferromagnetic features
in ferrimagnets has been demonstrated.
The DMRG technique has allowed us to have access to very low
temperatures.
In an attempt at going further down to even lower temperatures, we
develop the MSW theory for ferrimagnets \cite{Yama91} in the next
section.
We not only aim at obtaining the precise description of the
low-temperature behavior but are also interested in describing the
overall thermal behavior based on the spin-wave picture.

\section{Modified Spin-Wave Approach}\label{S:MSW}

   For years the conventional spin-wave theory
\cite{Hols98,Ande94,Kubo68} was plagued by the difficulty
that the zero-field magnetization diverges in low-dimensional
magnets.
The low-temperature series expansion within the theory only brings us
the leading term of the specific heat and nothing correct for the
susceptibility.
However, imposing a constraint on the magnetization,
Takahashi \cite{Taka33,Taka68} succeeded in describing the
low-temperature thermodynamics of low-dimensional ferromagnets.
His idea was further applied to several quantum antiferromagnets
\cite{Taka94,Hirs69,Reze89,Hida94} and its wide applicability was
established.
Recently two of the present authors \cite{Yama91} have introduced
this modified spin-wave theory into quantum ferrimagnets and
demonstrated that it is quite useful in understanding their
characteristic features.
Here we develop our argument taking into account interactions between
spin waves.

\subsection{Dispersion relations}

   We have observed that the low-temperature thermodynamics well
reflects the dispersion relation of the ferromagnetic branch,
whereas the intermediate-temperature behavior is well attributed to
that of the gapped antiferromagnetic branch.
This scenario is indeed supported by the MSW theory \cite{Yama91}.
A combination of the ferromagnetic and antiferromagnetic spin waves
on which a particular constraint is imposed reproduces the $T^{1/2}$
initial behavior, the Schottky-like peak, and the $T^{-2}$ decay of
the specific heat, and the $T^{-2}$ divergence and the $T^{-1}$ decay
of the susceptibility (see Fig. \ref{F:SW-C} and Fig. \ref{F:SW-S}).
However, we should be reminded that the conventional spin-wave
calculation, based on which we have started our first attempt
\cite{Yama91} to construct the thermodynamics, considerably
underestimates the antiferromagnetic gap ${\mit\Delta}$.
Consequently, the MSW approach succeeds in reproducing the
Schottky anomaly itself, to be sure, but fails in locating it at
a correct temperature.
In order to obtain a better description, we here refine the
original spin-wave theory.

   Let us introduce bosonic operators through the Holstein-Primakoff
transformation \cite{Hols98}
\begin{equation}
   \left.
   \begin{array}{ll}
      S_j^+=(2S-a_j^\dagger a_j)^{1/2}a_j\,,&
      S_j^z=S-a_j^\dagger a_j\,,\\
      s_j^+=b_j^\dagger(2s-b_j^\dagger b_j)^{1/2}\,,&
      s_j^z=-s+b_j^\dagger b_j\,,
   \end{array}
   \right.
   \label{E:HPBoson}
\end{equation}
where we regard $S$ and $s$ as quantities of the same order, that
is, $O(S)=O(s)$.
The Hamiltonian (\ref{E:H}) with $\delta=1$ and $H=0$ is expressed in
terms of the bosonic operators as
\begin{equation}
   {\cal H}=E_{\rm class}+{\cal H}_0+{\cal H}_1+O(S^{-1})\,,
   \label{E:Hboson}
\end{equation}
where
\begin{eqnarray}
   &&
   E_{\rm class}=-2sSJN\,,
   \label{E:Ecl}\\
   &&
   {\cal H}_0
     =J\sum_j
      \biggl\{
       2s a_j^\dagger a_j+2S b_j^\dagger b_j
   \nonumber \\
   &&\qquad\qquad
      +\sqrt{sS}
       \Bigl[
        (a_j+a_{j+1})b_j+\mbox{h.c.}
       \Bigr]
      \biggr\}\,,
   \label{E:H0} \\
   &&
   {\cal H}_1
     =-J\sum_j
      \biggl\{
       \frac{1}{4}
       \Bigl[
        \sqrt{S/s}\,(a_j+a_{j+1})b_j^\dagger b_j^2
   \nonumber \\
   &&\qquad\qquad\qquad
       +\sqrt{s/S}\,a_j^\dagger a_j^2(b_j+b_{j-1})
       +\mbox{h.c.}
       \Bigr]
   \nonumber \\
   &&\qquad\qquad\qquad
     +(a_j^\dagger a_j+a_{j+1}^\dagger a_{j+1})b_j^\dagger b_n
      \biggr\}\,.
   \label{E:H1}
\end{eqnarray}
The treatment of the quartic interaction (\ref{E:H1}) is not so
canonical as that of the quadratic Hamiltonian (\ref{E:H0}).
A variety of approaches are possible\cite{Ivan14}.
Here, in an attempt to obtain the dispersion relations beyond the
noninteracting spin-wave theory, we first diagonalize ${\cal H}_0$ and
next extract corrections from ${\cal H}_1$.

   The Bogoliubov transformation
\begin{equation}
   \left.
   \begin{array}{c}
      \alpha_k=\mbox{cosh}\theta_k\ a_k
              +\mbox{sinh}\theta_k\ b_k^\dagger\,,\\
      \beta_k =\mbox{sinh}\theta_k\ a_k^\dagger
              +\mbox{cosh}\theta_k\ b_k\,,
   \end{array}
   \right.
   \label{E:Bogliubov}
\end{equation}
combined with
\begin{equation}
   \left.
   \begin{array}{c}
      a_k={\displaystyle \frac{1}{\sqrt{N}}}
          {\displaystyle \sum_j} 
          {\rm e}^{ {\rm i}k(j-1/4)}a_j\,,\\
      b_k={\displaystyle \frac{1}{\sqrt{N}}}
          {\displaystyle \sum_j} 
          {\rm e}^{-{\rm i}k(j+1/4)}b_j\,,\\
   \end{array}
   \right.
   \label{E:Fourier}
\end{equation}
and 
\begin{equation}
   \mbox{tanh}2\theta_k
     =-\frac{2\sqrt{Ss}}{S+s}
      \mbox{cos}\,\biggl(\frac{k}{2}\biggl)\,,
   \label{E:theta}
\end{equation}
diagonalizes ${\cal H}_0$ as \cite{Pati94,Breh21}
\begin{equation}
   {\cal H}_0
     =E_0
     +J\sum_k
      \left(
       \omega_{k}^-\alpha_k^\dagger\alpha_k
      +\omega_{k}^+\beta_k^\dagger \beta_k
      \right)\,,
   \label{E:diagH0}
\end{equation}
where we have taken twice the lattice constant as unity.
The first term in (\ref{E:diagH0}) is a quantum correction to the
ground-state energy of order $O(S^1)$,
\begin{equation}
   E_0=J\sum_k
   \Bigl[
    \omega_k-(S+s)
   \Bigr]\,,
   \label{E:E0}
\end{equation}
with
\begin{equation}
   \omega_k=\sqrt{(S-s)^2+4Ss\sin^2(k/2)}\,.
\end{equation}
The following terms are the ferromagnetic and antiferromagnetic
spin-wave modes of order $O(S^1)$, whose dispersion relations are,
respectively, given by
\begin{equation}
   \omega_{k}^\mp=\omega_k\mp(S-s)\,.
\end{equation}
The lower-energy mode shows a quadratic dispersion relation at small
$k$'s, which is consistent with the $T^{1/2}$ initial
behavior of the specific heat.
On the other hand, the gap between the two branches is exactly $J$,
which considerably contradicts the numerical estimate $1.759J$.

   Now we pick up relevant contributions to the dispersions, as well
as to the ground-state energy, from ${\cal H}_1$.
Employing the Wick theorem, we rewrite ${\cal H}_1$ as
\begin{eqnarray}
   {\cal H}_1
   &=&E_1-J\sum_k
    \left(
    \delta\omega_k^-\alpha_k^\dagger\alpha_k
    +\delta\omega_k^+\beta_k^\dagger\beta_k
    \right)
   \nonumber\\
   &+&
     {\cal H}_{\rm irrel}+{\cal H}_{\rm quart}\,,
\end{eqnarray}
where $H_{\rm irrel}$ contains irrelevant terms such as $\alpha_k\beta_k$
and ${\cal H}_{\rm quart}$ contains residual two-body interactions, both of
which are neglected in the following.
The correction to the ground-state energy, $E_1$, and ones to the
dispersions, $\delta\omega_k^\pm$, are, respectively, given by
\begin{eqnarray}
   &&
   E_1=-2JN
       \left[
        {\mit\Gamma}_1^2+{\mit\Gamma}_2^2
       +\left(
         \sqrt{S/s}+\sqrt{s/S}
        \right)
        {\mit\Gamma}_1{\mit\Gamma}_2
       \right]\,,
   \label{E:E1} \\
   &&
   \delta\omega_k^\pm
      =2(S+s){\mit\Gamma}_1
       \frac{\sin^2\bigl(\frac{k}{2}\bigr)}{\omega_k}
      +\frac{{\mit\Gamma}_2}{\sqrt{Ss}}
       \Bigl[
        \omega_k\pm (S-s)
       \Bigr]\,,
    \label{E:domega}
\end{eqnarray}
with
\begin{equation}
   \left.
   \begin{array}{l}
      {\mit\Gamma}_1
        ={\displaystyle \frac{1}{2N}}
         {\displaystyle \sum_k}
         ({\rm cosh}2\theta_k-1)\,,\\
      {\mit\Gamma}_2
        ={\displaystyle \frac{1}{2N}}
         {\displaystyle \sum_k}
         {\displaystyle \cos\biggl(\frac{k}{2}\biggr)}\,
         {\rm sinh}2\theta_k\,.\\
   \end{array}
   \right.
   \label{E:Gamma}
\end{equation}
In the thermodynamic limit, the key constants ${\mit\Gamma}_1$ and
${\mit\Gamma}_2$ with $(S,s)=(1,1/2)$ are estimated to be
$0.304887$ and $-0.337779$, respectively.
Up to order $O(S^0)$, we end up with the Hamiltonian
\begin{equation}
   {\cal H}
     \simeq E_{\rm g}
    +J\sum_k
     \left(
      {\widetilde\omega}_k^- \alpha_k^\dagger \alpha_k
     +{\widetilde\omega}_k^+ \beta_k^\dagger  \beta_k
     \right)\,,
\end{equation}
where
\begin{eqnarray}
   &&
   {\widetilde\omega}_k^\pm=\omega_k^\pm-\delta\omega_k^\pm\,,
   \\
   &&
   E_{\rm g}=E_{\rm class}+E_0+E_1\,.
\end{eqnarray}

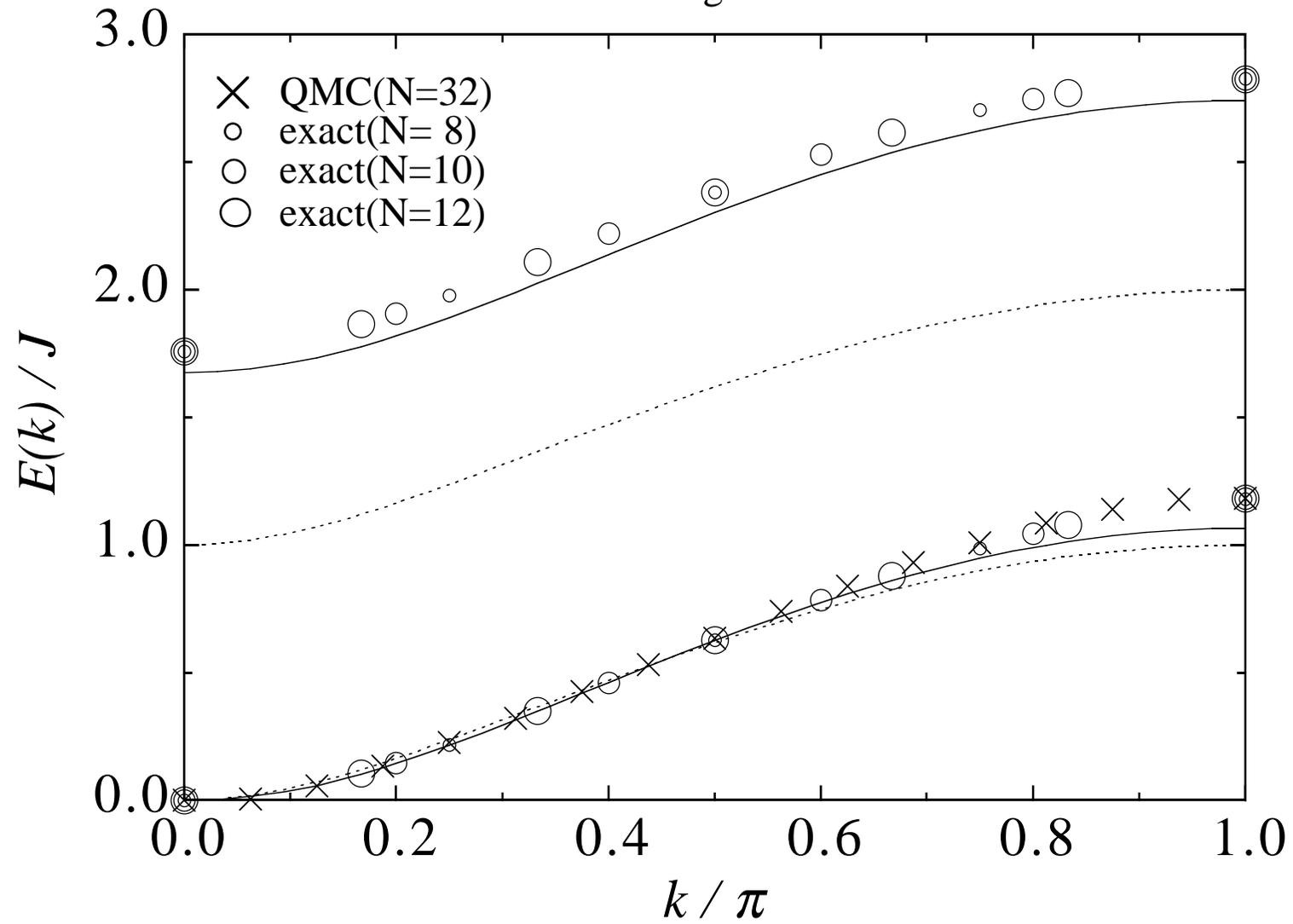
\begin{figure}
\caption{Dispersion relations of the lowest-energy states in the
         subspaces of $M=N/2\mp 1$:
         The noninteracting spin-wave result (dotted lines) and an
         improved calculation taking into account interactions between
         spin waves (solid lines).
         Previous numerical calculations [14] are also shown for the
         sake of comparison.
         Here we plot the excitation energy $E(k)$ taking the
         ground-state energy and twice the lattice constant as zero
         and unity, respectively.}
\label{F:Disp}
\end{figure}

   In Fig. \ref{F:Disp} we plot ${\widetilde\omega}_k^\pm$ and
$\omega_k^\pm$ as functions of $k$ at $(S,s)=(1,1/2)$ with the
previous numerical estimates \cite{Yama09} for finite chains.
We find that the antiferromagnetic mode is now improved to a great
extent.
Furthermore the band widths of the two branches, which are exactly
the same within the noninteracting spin-wave theory, are now indeed
different, due to the interactions.
The gap ${\mit\Delta}=J$ is replaced by
${\mit\Delta}=(1-2{\mit\Gamma}_2)J\simeq 1.676J$, which is much closer
to the exact value $1.759J$.
The ground-state energy $E_{\rm g}$ is also refined:
$E_{\rm class}+E_0\simeq -1.437J$, while
$E_{\rm class}+E_0+E_1\simeq -1.459J$, where the exact value is
$-1.454J$.

\subsection{Thermodynamics}

   Now let us start our MSW theory from the above-obtained improved
dispersion relations.
At finite temperatures we replace $\alpha_k^\dagger\alpha_k$ and
$\beta_k^\dagger\beta_k$ in the spin-wave Hamiltonian by
$\widetilde n^\pm_k\equiv\sum_{n^-,n^+}n^\pm P_k(n^-,n^+)$,
where
$P_k(n^-,n^+)$ is the probability of $n^-$ ferromagnetic and 
$n^+$ antiferromagnetic spin waves appearing in the $k$-momentum 
state and
satisfies
\begin{equation}
   \sum_{n^-,n^+} P_k(n^-,n^+)=1\,,
   \label{E:constP}
\end{equation}
for all $k$'s.
Then the free energy at zero field is given by
\begin{eqnarray}
   F&=&E_{\rm g}
      +\sum_k
        (\widetilde n^-_k\omega_k^-+\widetilde n^+_k\omega_k^+)
         \nonumber\\                                                  
    &+&k_{\rm B}T
      \sum_k\sum_{n^-,n^+}P_k(n^-,n^+){\rm ln}P_k(n^-,n^+)\,.
   \label{E:F}
\end{eqnarray}
We now carry out the minimization of the free energy (\ref{E:F}) with
respect to $P_k(n^-,n^+)$'s under a particular constraint
as well as the trivial constraint (\ref{E:constP}).
The original idea introduced by Takahashi \cite{Taka33,Taka68}
was that the zero-field magnetization should be zero.
This constraint works quite well especially in ferromagnets, where
it serves to control the number of Holstein-Primakoff bossons.
Let us apply the same constraint to the present
model:
\begin{equation}
   \langle S^z+s^z\rangle
     =(S-s)N
     -\sum_k\sum_{\sigma=\pm}
      \sigma{\widetilde n}_k^{-\sigma}
     =0\,,
   \label{E:constM}
\end{equation}
where $S^z=\sum_j S_j^z$ and $s^z=\sum_j s_j^z$.
Equation (\ref{E:constM}) claims that the thermal fluctuation
$\sum_k\sum_\sigma\sigma{\widetilde n}_k^{-\sigma}$ should be
constrained to take the {\it classical} magnetization $(S-s)N$.
The free energy and the magnetic susceptibility at thermal
equilibrium are then given by
\begin{eqnarray}
   &&F=E_{\rm g}
      +\mu(S-s)N
      -k_{\rm B}T\sum_{k}\sum_{\sigma=\pm}
       {\rm ln}(1+\widetilde n^\sigma_k)
     \,,\label{E:M-F}\\
   &&\chi=\frac{(g\mu_{\rm B})^2}{3k_{\rm B}T}
          \sum_k\sum_{\sigma=\pm}
          \widetilde n^\sigma_k(1+\widetilde n^\sigma_k)
     \,,\label{E:M-S}
\end{eqnarray}
with
\begin{equation}
   {\widetilde n}_k^\pm
     =\frac{1}
      {{\mbox e}^{({\widetilde\omega}_k^\pm \pm\mu)/k_{\rm B}T}-1}
     \,,
   \label{E:M-nk}
\end{equation}
where $\mu$ is a Lagrange multiplier due to the condition
(\ref{E:constM}).
The susceptibility has been obtained by calculating 
$\chi=(g\mu_{\rm B})^2(\langle M^2\rangle-\langle M\rangle^2)/3NT$ 
\cite{Taka33}.
Equations (\ref{E:M-F}) and (\ref{E:M-S}) are expanded in powers of
$T^{1/2}$ at low temperatures as
\widetext                                                             
\begin{eqnarray}
   &&
   \frac{C}{Nk_{\rm B}}
      = \frac{3}{4}\left(\frac{S-s}{Ss}\right)^{\frac{1}{2}}
         \frac{\zeta(\frac{3}{2})}{\sqrt{2\pi}}\widetilde t^{\frac{1}{2}}
      -\frac{1}{Ss}\widetilde t 
      + \frac{15}{32(S-s)^{\frac{1}{2}}(Ss)^{\frac{3}{2}}}
         \left[
          \frac{(S^2+Ss+s^2)\zeta(\frac{5}{2})}
              {\sqrt{2\pi}}
         -\frac{4\zeta(\frac{1}{2})}{\sqrt{2\pi}}
         \right]
        \widetilde t^{\frac{3}{2}}
        +O(\widetilde t^2)
         \,,\label{E:LTSE-C}\\
   &&
   \frac{\chi J}{N(g\mu_{\rm B})^2}
      = \frac{Ss(S-s)^2}{3}\widetilde t^{-2}
      -(Ss)^{\frac{1}{2}}(S-s)^{\frac{3}{2}}
       \frac{\zeta(\frac{1}{2})}{\sqrt{2\pi}}\widetilde t^{-\frac{3}{2}}
      +(S-s)\left[\frac{\zeta(\frac{1}{2})}{\sqrt{2\pi}}\right]^2
       \widetilde t^{-1}
      +O(\widetilde t^{-\frac{1}{2}})
         \,,\label{E:LTSE-S}
\end{eqnarray}
\narrowtext
where $\zeta(z)$ is Riemann's zeta function and
${\widetilde t}=k_{\rm B}{\widetilde T}/J=k_{\rm B}T/J\gamma$ with 
$\gamma=1-{\mit\Gamma}_1(S+s)/Ss-{\mit\Gamma}_2/Ss$.
Surprisingly, the low-temperature series expansions (\ref{E:LTSE-C})
and (\ref{E:LTSE-S}) are exactly the same as the thermodynamic
Bethe-ansatz calculations for the spin-$1/2$ ferromagnet
\cite{Taka08} except for $\gamma$.
Thus we recognize similarities between ferrimagnets and ferromagnets
at low temperatures.
We note, however, that ferrimagnets of $S=2s$ should not strictly be
identified with spin-$s$ ferromagnets because of the scaling factor
$\gamma$.
With the interactions, the original temperature $T$ is replaced by
${\widetilde T}(>T)$ or equivalently the original spins are reduced.
Therefore ferrimagnets of $S=2s$ behaves like
spin-${\widetilde s}(<s)$ ferromagnets, at least at low temperatures,
which reminds us of the quantum spin reduction \cite{Breh21} in
ferrimagnets.
The spin-wave theory shows that the staggered magnetization is
reduced to $(S+s)N-2\tau$ with
$2\tau=\sum_k[(S+s)/s\omega_k-1]$.

   Unfortunately, in the present model, the 
constraint introduced above is not useful at all at high temperatures 
because it allows
the number of bosons of each mode to diverge.
Actually, under the condition of zero magnetization, the specific
heat becomes a monotonically increasing function.
Hence we propose an alternative constraint \cite{Yama91}.
Let us consider the minimization of the free energy constraining
the {\it staggered magnetization} to be zero:
\begin{equation}
   \langle :S^z-s^z: \rangle
      =(S+s)N-(S+s)\sum_k\sum_{\sigma=\pm}
       \frac{\widetilde n^\sigma_k}{\omega_k}
      =0\,,
   \label{E:constMst}
\end{equation}
where the normal ordering is taken with respect to both operators
$\alpha$ and $\beta$.
Equation (\ref{E:constMst}) implies, just as Eq. (\ref{E:constM})
does, that the thermal fluctuation
$(S+s)\sum_k\sum_\sigma{\widetilde n}_k^\sigma/\omega_k$ should take
the {\it classical} value $(S+s)N$, rather than the renormalized value
$(S+s)N-2\tau$.
We note that, based on the naive idea of zero staggered
magnetization without the normal ordering, we have
\begin{equation}
   \langle S^z-s^z \rangle
      =(S+s)N-2\tau-(S+s)\sum_k\sum_{\sigma=\pm}
       \frac{\widetilde n^\sigma_k}{\omega_k}
      =0\,.
   \label{E:constMstx}
\end{equation}
Besides such a phenomenological argument, a more stringent reason
drives us to adopt Eq. (\ref{E:constMst}):
The condition (\ref{E:constMst}) exactly reproduces the
low-temperature series expansions (\ref{E:LTSE-C}) and
(\ref{E:LTSE-S}), whereas the condition (\ref{E:constMstx}) fails
to do.
The conventional spin-wave theory at least brings the correct
leading term of the specific heat, which is obviously reproduced by
the MSW theory with the constraint of zero magnetization.
The low-temperature behavior must be inherent in the dispersion
relation and should not be influenced by supplementary constraints.
Now the set of self-consistent equations (\ref{E:M-F}) and (\ref{E:M-S})
is replaced by
\begin{eqnarray}
   &&F=E_{\rm g}
      +\mu(S+s)N
      -k_{\rm B}T\sum_{k}\sum_{\sigma=\pm}
       {\rm ln}(1+\widetilde n^\sigma_k)
     \,,\label{E:Mst-F}\\
   &&\chi=\frac{(g\mu_{\rm B})^2}{3k_{\rm B}T}
          \sum_k\sum_{\sigma=\pm}
          \widetilde n^\sigma_k(1+\widetilde n^\sigma_k)
     \,,\label{E:Mst-S}
\end{eqnarray}
with 
\begin{equation}
   {\widetilde n}_k^\pm
     =\frac{1}
      {{\mbox e}^{[J{\widetilde\omega}_k^\pm
      -\mu(S+s)/\omega_k]/k_{\rm B}T}-1}
     \,,
   \label{E:Mst-nk}
\end{equation}
where $\mu$ is a Lagrange multiplier due to the condition
(\ref{E:constMst}).

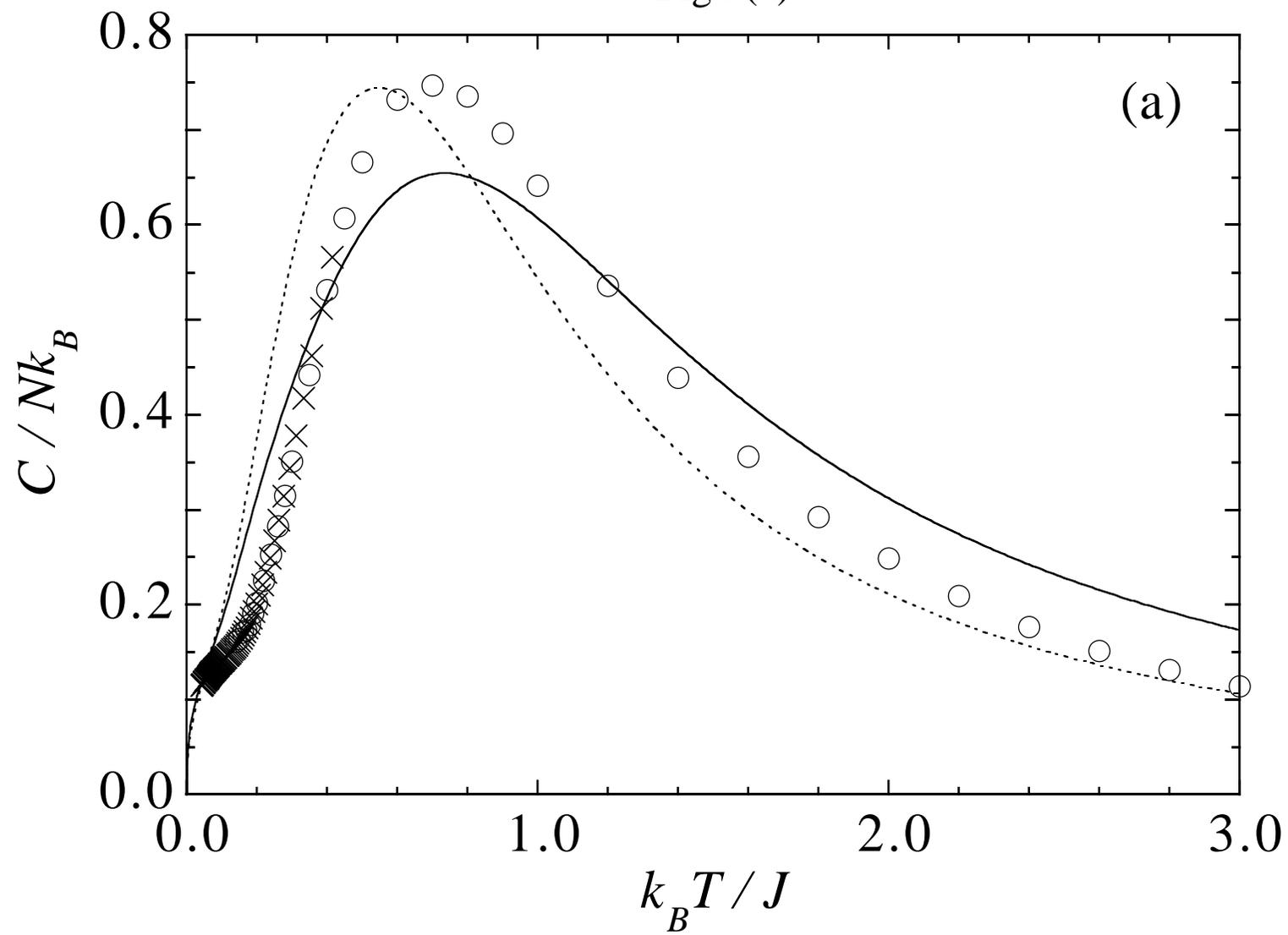
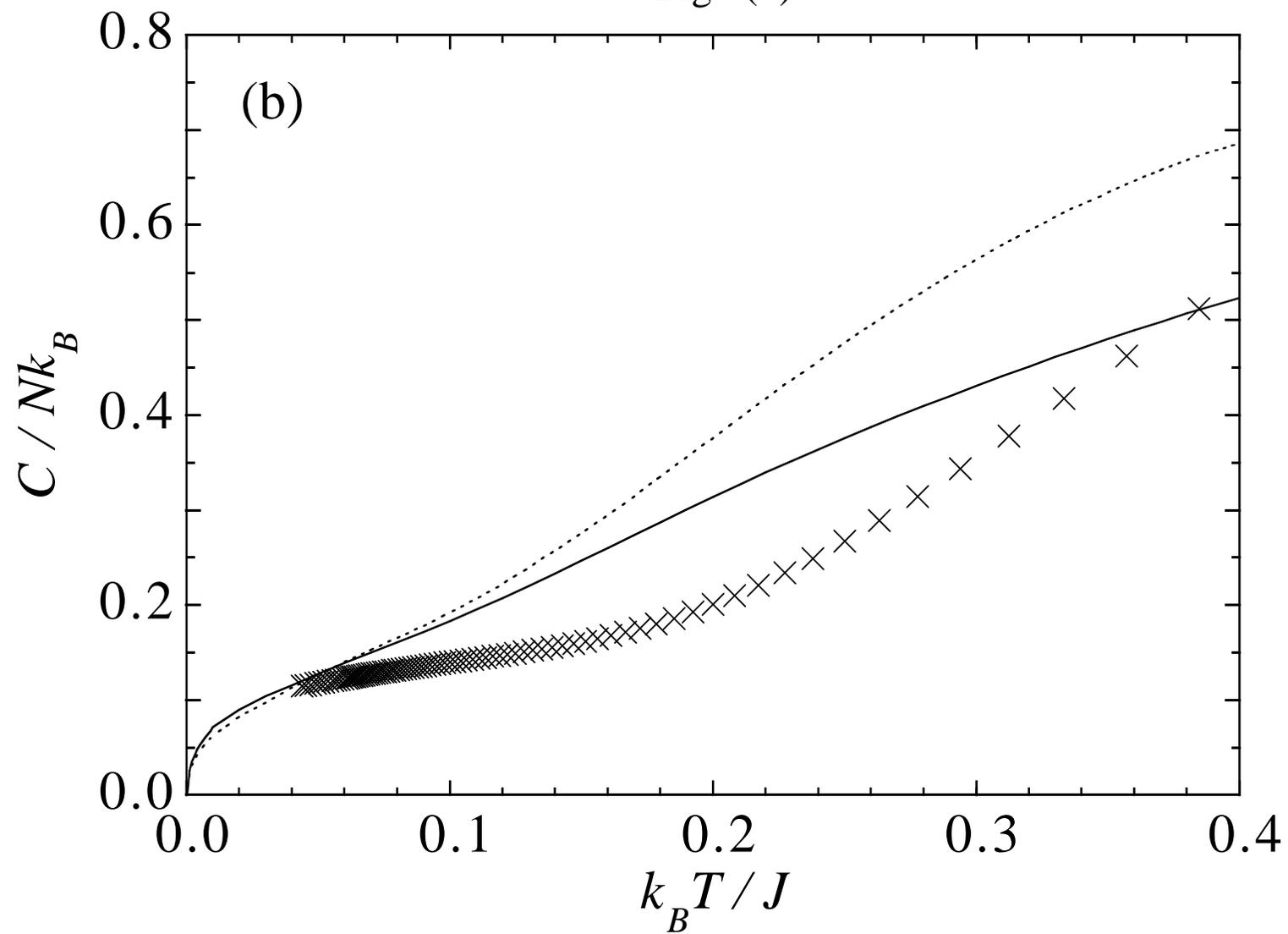
\begin{figure}
\caption{Temperature dependence of the specific heat per unit cell:
         The MSW calculation with noninteracting spin waves (dotted
         lines) and that with the improved dispersion relations (solid
         lines).
         Numerical findings [QMC ($\bigcirc$) and DMRG($\times$)] are
         also shown.}
\label{F:SW-C}
\end{figure}

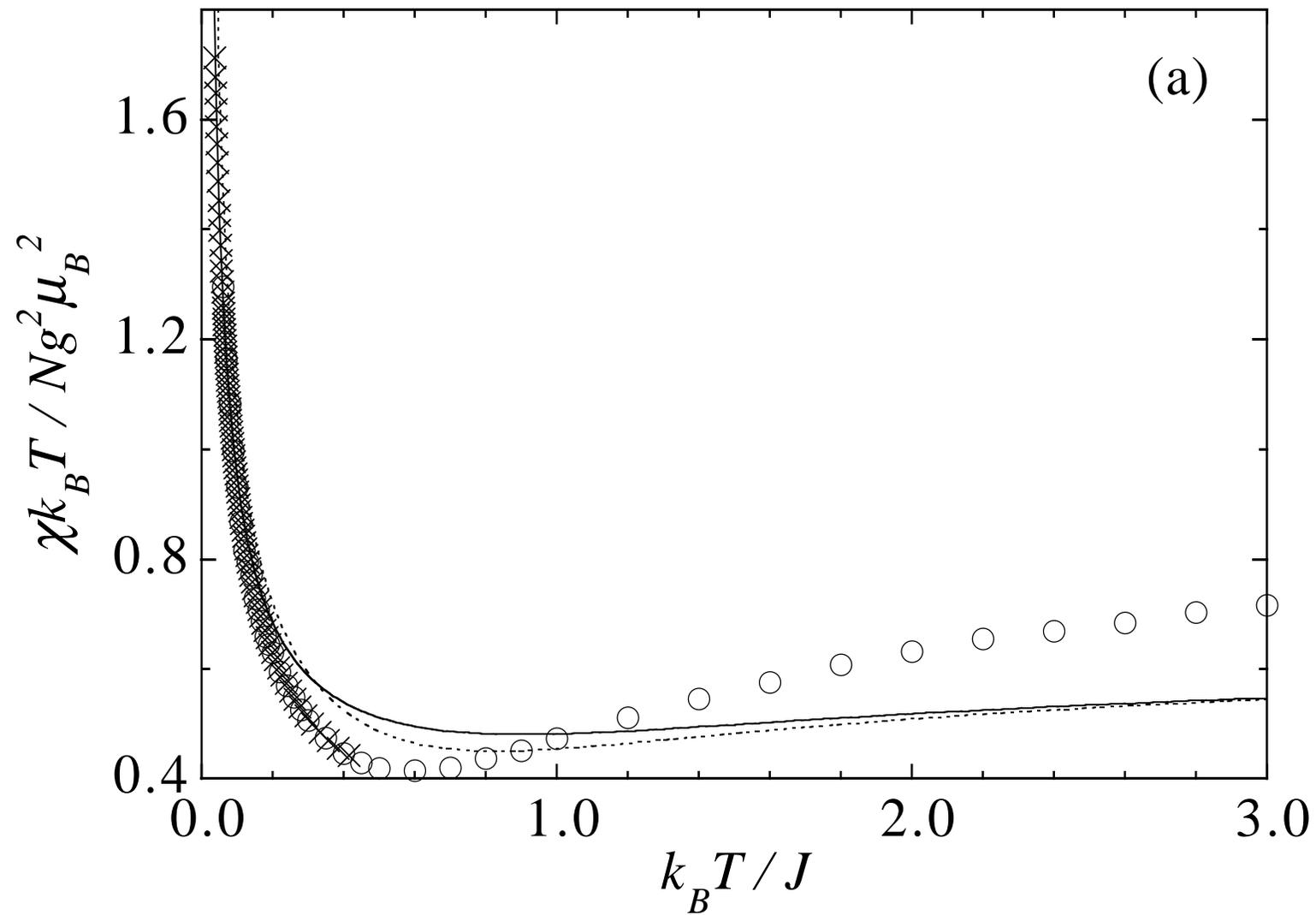
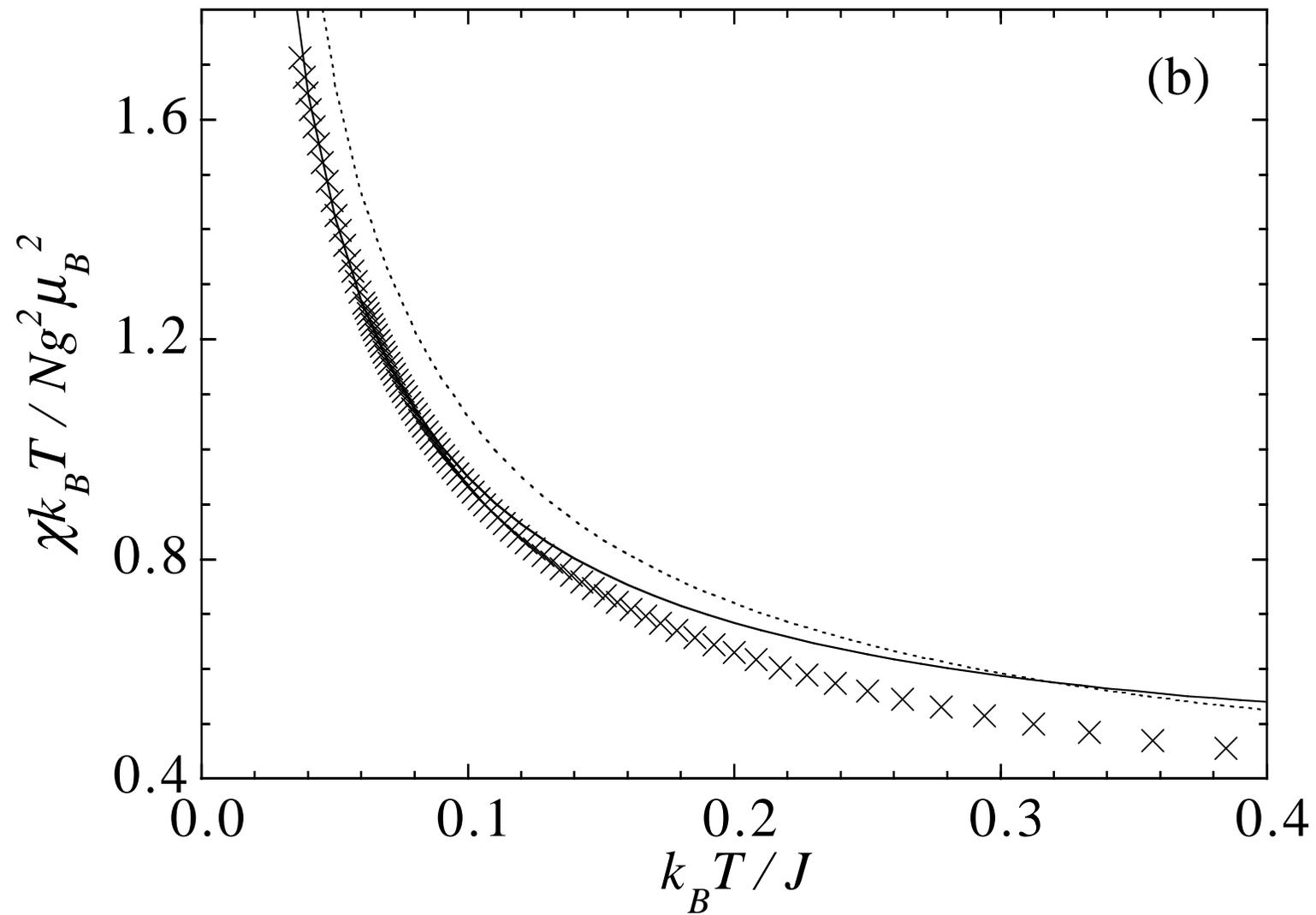
\begin{figure}
\caption{Temperature dependence of the magnetic susceptibility times
         temperature per unit cell:
         The MSW calculation with noninteracting spin waves (dotted
         lines) and that with the improved dispersion relations (solid
         lines).
         Numerical findings [QMC ($\bigcirc$) and DMRG($\times$)] are
         also shown.}
\label{F:SW-S}
\end{figure}

   In the case of $(S,s)=(1,1/2)$, we have numerically obtained
(\ref{E:Mst-F}) and (\ref{E:Mst-S}) in the thermodynamic limit and
visualize them in Fig. \ref{F:SW-C} and Fig. \ref{F:SW-S}, where the
solid curves represent the calculations starting from the improved
dispersion relations ${\widetilde\omega}_k^\pm$, while the dashed
curves the calculations with $\omega_k^\pm$ \cite{Yama91} instead of
${\widetilde\omega}_k^\pm$.
Though the present calculation including the $O(S^0)$ interactions
underestimates the height of the Schottky-like peak,
the interactions indeed correct the location of the peak, which
emphasizes that the gapped antiferromagnetic spin-wave mode
predominantly contributes to this peak.
We would like to attribute the discrepancy unsolved to the
constraint rather than to the dispersion relations.
Even though we adopt the two constraints (\ref{E:constM}) and
(\ref{E:constMst}) simultaneously, the results do not change
essentially.
This is not so surprising because the two constraints play almost
the same role at low temperatures, as the low-temperature series
expansions imply, whereas only the constraint (\ref{E:constMst}) is
relevant at high temperatures.
The precise description of the overall temperature dependence might
be obtained with a temperature-dependent constraint, which is not
so interesting.

   We obtain the best results from the MSW theory at low temperatures.
Figure \ref{F:SW-S}(b) fully convinces us of the validity of the
present MSW calculation.
Figure \ref{F:SW-C}(b) clearly reveals the low-temperature behavior
which no numerical tool has succeeded in observing.
Takahashi compared his MSW findings \cite{Taka68} for ferromagnets
with the spin-$1/2$ thermodynamic Bethe-ansatz calculations
\cite{Taka08} and found that the MSW theory correctly describes the
leading two terms of the specific heat and the leading three terms
of the susceptibility in its low-temperature series expansions.
We also find that the low-temperature series expansions
(\ref{E:LTSE-C}) and (\ref{E:LTSE-S}) coincide with ones of the
spin-$1/2$ ferromagnet to the same extent except for the scaling
factor $\gamma$.
That may be why the MSW and numerical findings for the
susceptibility join at higher temperatures than those for the specific
heat.

\section{Summary}\label{S:S}

   Developing an analytic argument as well as employing various
numerical tools, we have investigated the thermodynamic properties of
Heisenberg ferrimagnetic spin chains.
Both ferromagnetic and antiferromagnetic aspects lie in the model;
they are most clearly exhibited at low and intermediate
temperatures, respectively.
One might say that mixed-spin chains possess mixed features.
We have shown that the MSW theory starting from the improved
dispersion relations precisely describes the low-temperature
behavior of the model.
We appeal to experimentalists to carry out specific heat and susceptibility
measurements of mixed-spin materials especially at low temperatures.
On the other hand, the Schottky-like peak of the specific heat and
the minimum of the susceptibility-temperature product clearly
reflecting the antiferromagnetic gap imply that the
antiferromagnetic excitations are not smeared out in the ferromagnetic
spectra but stand out clearly.
Hence neutron-scattering measurements are also encouraged.

\acknowledgments

   The authors would like to thank H.-J. Mikeska, S. Brehmer, and
S. K. Pati for their useful comments and fruitful discussions.
This work was supported by the Japanese Ministry of Education, Science,
and Culture through the Grant-in-Aid No. 09740286 and by a Grant-in-Aid
from the Okayama Foundation for Science and Technology.
T.F. is supported by JSPS Postdoctoral Fellowship for Research Abroad.
Part of the numerical computation was done using the facility of the
Supercomputer Center, Institute for Solid State Physics, University of
Tokyo.

\begin{figure}
\caption{Schematic representations of
         the $M=N/2$ ground state of the $N=1$ isolated dimer composed
         of spin $1$ and spin $1/2$ and its ferromagnetic (b) and
         antiferromagnetic (c) excitations.
         The arrow (the bullet symbol) denotes a spin $1/2$ with its
         fixed (unfixed) projection value.
         The solid (broken) segment is a singlet (triplet) pair.
         The circle represents an operation of constructing a spin $1$ by
         symmetrizing the two spin $1/2$'s inside.}
\label{F:dimer}
\end{figure}

\widetext
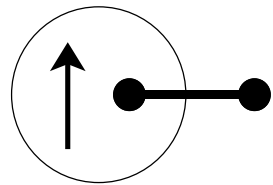
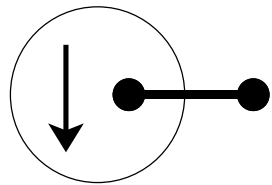
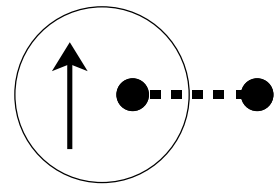
\end{document}